\documentclass{article}

\usepackage[preprint]{infocog_neurips_2022}

\usepackage[utf8]{inputenc} 
\usepackage[T1]{fontenc}    
\usepackage{hyperref}       
\usepackage{url}            
\usepackage{booktabs}       
\usepackage{amsfonts}       
\usepackage{nicefrac}       
\usepackage{microtype}      
\usepackage{xcolor}         

\usepackage{amsmath}
\usepackage{amsthm}
\usepackage{amssymb}

\newcommand{\kwad}[1]{\left[ #1 \right]}
\newcommand{\klam}[1]{\left\{ #1 \right\}}
\newcommand{\boole}[1]{{\bf 1}{\klam{#1}}}

\DeclareMathOperator{\card}{\#}
\DeclareMathOperator{\mean}{\mathbf{E}}
\DeclareMathOperator*{\hilberg}{hilb}

\title{There Are Fewer Facts Than Words: \\ Communication With A
  Growing Complexity}

\author{%
  Łukasz Dębowski
  \\
  Institute of Computer Science\\
  Polish Academy of Sciences\\
  01-248 Warszawa, Poland\\
  \texttt{ldebowsk@ipipan.waw.pl} \\
}

\begin{document}

\maketitle


\begin{abstract}
  We present an impossibility result, called a theorem about facts and
  words, which pertains to a general communication system. The theorem
  states that the number of distinct words used in a finite text is
  roughly greater than the number of independent elementary persistent
  facts described in the same text. In particular, this theorem can be
  related to Zipf's law, power-law scaling of mutual information, and
  power-law-tailed learning curves. The assumptions of the theorem
  are: a finite alphabet, linear sequence of symbols, complexity that
  does not decrease in time, entropy rate that can be estimated, and
  finiteness of the inverse complexity rate.
\end{abstract}

\section{Introduction}

In several recent large-scale computational experiments in statistical
language modeling, there were observed power-law tails of learning
curves \citep{TakahiraOthers16, HestnessOther17, HahnFutrell19,
  BravermanOther19, KaplanOther20, HenighanOther2020,
  HernandezOther21, TanakaIshii21}. Namely, the difference between the
cross entropy rate of the statistical language model and the entropy
rate of natural language decays as a power law with respect to the
amount of training data. Equivalently, this is tantamount to a
power-law growth of mutual information between increasing blocks of
text---the first observation thereof attributed to \citet{Hilberg90},
see also \citep{CrutchfieldFeldman03}. This power-law growth occurs
for languages typologically as diverse as English, French, Russian,
Chinese, Korean, and Japanese.  Moreover, we observe a universal
language-independent value of the power-law exponent: the mutual
information between two blocks of length $n$ is proportional to
$n^{0.8}$ \citep{TakahiraOthers16, TanakaIshii21}.

In this paper, we would like to advertise some mathematical theory of
this phenomenon, covering its potential causes and effects. We have
been developing this theory for several years. Our results were
resumed in the recently published book \citep{Debowski21} and the
subsequent article \citep{Debowski21b}. This paper supplies an
updated brief overview for a venue of machine learning. The novel
thing is a simple generalization to non-stationary processes with a
growing complexity.

The basic theory of power-law-tailed learning curves can be simply
stated as furnishing the proof of a general statement of form:
\begin{quote}
  \begin{center}
    \bfseries The number of distinct words used in a finite text is
    roughly greater than \\ the number of independent elementary
    persistent facts described in this text.
\end{center}
\end{quote}
We call this sort of a statement a theorem about facts and words.  In
fact, the theorems about facts and words come into a few distinct
flavors and can be proved relatively easily provided a certain
attention is paid to the formal understanding of the concepts of a
fact and of a word.

The theorems about facts and words can be regarded as an impossibility
result that pertains to a general communication system. Simply
speaking, one cannot communicate about a certain amount of independent
facts in a repetitive fashion without effectively using at least as
many distinct words. This result seems paradoxical since we might
think that combining words we may express many more independent facts.
Moreover, the above result links information theory and discrete
stochastic processes with linguistics, semantics, and cognition. In
principle, it applies to any kind of a communication system consisting
of a finite number of discrete signs stringed into longer
messages. Besides natural language, some obvious examples are computer
programs, DNA, and music.

In particular, the statements and the proofs of the theorems about facts
and words combine:
\begin{itemize}
\item Zipf's and Herdan-Heaps' laws for word frequency
  distributions \citep{Zipf35, Mandelbrot54, Guiraud54, Herdan64,
    Heaps78},
\item universal coding based on grammars \citep{DeMarcken96,
    KiefferYang00, CharikarOthers05} and on normalized maximum
  likelihood \citep{Shtarkov87en2, Ryabko88en2, Ryabko08},
\item consistent (hidden) Markov order estimators
  \citep{MerhavGutmanZiv89, ZivMerhav92},
\item the concept of infinite excess entropy \citep{Hilberg90,
    EbelingNicolis91, EbelingPoschel94, BialekNemenmanTishby01b,
    CrutchfieldFeldman03},
\item the ergodic theorem and the ergodic decomposition
  \citep{Birkhoff32, Rohlin49en2, GrayDavisson74}, and
\item Kolmogorov complexity and algorithmic randomness \citep{Kolmogorov65en2,
    MartinLof66, LiVitanyi08}.
\end{itemize}
As we can see, there are many interacting mathematical concepts. There
are also many open problems in the surrounding theory. In the
following, we present some particular version of a theorem about facts
and words, which pertains to algorithmic randomness \citep{Debowski21}
and consistent Markov order estimation \citep{Debowski21b}. An
informal discussion is relegated to Appendix \ref{secDiscussion}.

\section{Preliminaries}

Consider a string $x_j^k:=(x_j,x_{j+1},...,x_k)$ over a countable
alphabet. Its prefix-free Kolmogorov complexity is denoted $K(x_j^k)$
\citep{LiVitanyi08}.  The algorithmic mutual information between
strings $u$ and $v$ is $J(u,v):=K(u)+K(v)-K(u,v)$. The expectation of
random variable $X$ is denoted $\mean X$. The Shannon entropy of $X$
is $H(X):=\mean\kwad{-\log P(X)}$ \citep{CoverThomas06}.

As in \citep[Definition 8.1]{Debowski21}, we will use the Hilberg
exponent of a sequence, defined as
\begin{align}
  \hilberg_{n\to\infty} S(n):=
  \kwad{\limsup_{n\to\infty} \frac{\log S(n)}{\log n}}_+,
  \quad
  r_+:=r\,\boole{r\ge 0}
  ,
\end{align}
so that $\hilberg_{n\to\infty} n^\beta=\beta$ for $\beta\ge 0$. We
recall that if limit $s=\lim_{n\to\infty} S(n)/n$ exists then
\begin{align}
  \hilberg_{n\to\infty} \kwad{S(n)-sn}\le\hilberg_{n\to\infty} \kwad{2S(n)-S(2n)}.
\end{align}

For a discrete one-sided stochastic process $(X_i)_{i\in\mathbb{N}}$, we
consider conditions:
\begin{itemize}
\item[(A)] The complexity rate
  $h=\displaystyle\lim_{n\to\infty} \mean K(X_1^n)/n$ exists and
  $\displaystyle\hilberg_{n\to\infty} \kwad{hn-H(X_1^n)}=0$.
\item[(B)] The complexity does not decrease in time:
  $\mean K(X_1^n)\le \mean K(X_{n+1}^{2n})$.
\item[(C)] The inverse complexity rate is finite,
  $H:=\displaystyle\limsup_{n\to\infty} \mean
  \frac{n}{K(X_1^n)}<\infty$ (thus $h>0$ for (A)).
\item [(D)] The alphabet is finite: $X_i:\Omega\to\klam{a_1,a_2,...,a_D}$,
  where $D\in\mathbb{N}$.
\end{itemize}
In particular, conditions (A) and (B) are satisfied by any stationary
process that satisfies (D). 

For conditions (A) and (B), we also obtain that the power-law rate of
redundancy is dominated by the power-law rate of mutual information,
\begin{align}
  \label{CompMI}
  \hilberg_{n\to\infty} \kwad{\mean K(X_1^n)-hn}\le
  \hilberg_{n\to\infty} \mean J(X_1^n;X_{n+1}^n).
\end{align}
Condition
$\displaystyle\hilberg_{n\to\infty} \mean J(X_1^n;X_{n+1}^n)>0$ is
called the Hilberg condition, after \citet{Hilberg90}.



\section{Facts}

The concept of independent elementary persistent facts can be most easily
understood on the example of a certain stationary ergodic process over
a countably infinite alphabet called a Santa Fe process
\citep{Debowski11b}. Let $(K_i)_{i\in\mathbb{N}}$ be an IID process
in natural numbers with the Zipfian marginal distribution
\begin{align}
  \label{Zipf}
  P(K_i=k)=\frac{k^{-\alpha}}{\zeta(\alpha)},
\end{align}
where $k\in\mathbb{N}$, $\alpha>1$ is a fixed parameter, and
$\zeta(\alpha):=\sum_{k=1}^\infty k^{-\alpha}$ is the Riemann zeta
function. Distribution (\ref{Zipf}) is a formal model of Zipf's law
from quantitative linguistics \citep{Zipf35, Mandelbrot54}. Moreover,
let $(z_k)_{k\in\mathbb{N}}$ be an algorithmically random sequence,
i.e., a sequence of particular (= fixed) bits (= coin flips) such that
the Kolmogorov complexity of any string $z_1^k$ is the highest
possible, $K(z_1^k)\ge k-c$ for a certain constant $c<\infty$ and all
lengths $k\in\mathbb{N}$ \citep{LiVitanyi08}.  Then the Santa Fe
process $(X_i)_{i\in\mathbb{N}}$ is a sequence of pairs
\begin{align}
  X_i=(K_i,z_{K_i}).
\end{align}
In the following, bits $z_k$ will be called facts.

The Santa Fe process can be understood as a model of an infinite text
that consists of random statements of form ,,the $k$-th fact equals
$z_k$''. Importantly, these statements are non-contradictory, namely,
if statements $X_i$ and $X_j$ describe the same fact ($K_i=K_j$) then
they assert the same value of this fact ($z_{K_i}=z_{K_j}$). Moreover,
we observe that facts $z_k$ are in some sense independent (the
Kolmogorov complexity of their concatenation is the highest possible),
elementary (they assume only two distinct values), and persistent
(described faithfully at any time instant $i$).

We will say that a finite text $x_1^n$ describes exactly first $l$
facts of a fixed sequence $(z_k)_{k\in\mathbb{N}}$ by means of a
computable function $g$ if $l=U_g(x_1^n;z_1^\infty)-1$, where
\begin{align}
  U_g(x_1^n;z_1^\infty):=\min
  \klam{k\in\mathbb{N}:g(k,x_1^n)\neq z_k}.
\end{align}
For a Santa Fe process we can easily construct a function $g$ such
that $U_g(X_1^n;z_1^\infty)=\min\klam{k\in\mathbb{N}: k\not\in K_1^n}$
by reading off the values of $z_k$ for all $k\in K_1^n$.  Hence it can
be proved that the expected number of initial facts described by a
random text $X_1^n$ grows as a power law
\begin{align}
  \label{Facts}
  \hilberg_{n\to\infty} \mean U_g(X_1^n;z_1^\infty)=1/\alpha\in(0,1).
\end{align}
Power laws (\ref{Zipf}) and (\ref{Facts}) are related to each other as
Zipf's law is related to Herdan-Heaps' law \citep{Guiraud54, Herdan64,
  Heaps78}. Processes $(X_i)_{i\in\mathbb{N}}$ such that
$\displaystyle\hilberg_{n\to\infty} \mean U_g(X_1^n;z_1^\infty)>0$ for
a certain algorithmically random sequence $(z_k)_{k\in\mathbb{N}}$ and
a certain computable function $g$ are called perigraphic. All
perigraphic processes have incomputable probability distributions. We
notice that Santa Fe processes are perigraphic.

These results can be linked to the redundancy rate. As shown in
\citep[Eqs. (8.117)--(8.119)]{Debowski21}, for any discrete stochastic
process $(X_i)_{i\in\mathbb{N}}$, any algorithmically random sequence
$(z_k)_{k\in\mathbb{N}}$, and any computable function $g$, we obtain
inequality
\begin{align}
  \label{PreFactsComp}
  \mean U_g(X_1^n;z_1^\infty)-6\log
  \mean U_g(X_1^n;z_1^\infty) -c_g
  \le
  \sup_{k\in\mathbb{N}} \mean J(X_1^n;z_1^k)
  \le
  \mean K(X_1^n)-H(X_1^n)
  .
\end{align}
where constant $c_g<\infty$ depends on function $g$. Hence, the
power-law rate of the number of facts is dominated by the power-law
rate of redundancy,
\begin{align}
  \label{FactsComp}
  \hilberg_{n\to\infty} \mean U_g(X_1^n;z_1^\infty)
  \le
  \hilberg_{n\to\infty} \kwad{\mean K(X_1^n)-hn}
\end{align}
if condition (A) holds. The above inequality can be chained with
inequality (\ref{CompMI}).

\section{Words}

In the remaining move, we will relate the algorithmic mutual
information to the number of words used in a given text. This can be
done in several ways. One way, pursued by \citet{Debowski11b}, is to
apply the minimal grammar-based coding
\citep{KiefferYang00,CharikarOthers05}, which applies a mathematical
concept of a word that resembles words in a linguistic sense
\citep{DeMarcken96} but this method is a bit lengthy to describe
formally.  Therefore, here, we will apply a different approach which
is based on consistent Markov order estimation
\citep{MerhavGutmanZiv89, ZivMerhav92}---that approach was pursued by
\citet{Debowski21b}.

There is a function called subword complexity that counts how many
distinct substrings of a given length there are in a given string,
namely,
\begin{align}
  V(k|x_1^n):=\card\klam{x_{i+1}^{i+k}:0\le i\le n-k}.
\end{align}
Function $V(k|x_1^n)$ will be a proxy for the number of distinct words
in text $x_1^n$. The only problem is to choose a motivated length $k$
of the substrings. In principle, this $k$ may depend on string
$x_1^n$.

Quite a natural choice of $k$ is the estimator of the Markov order of
the process defined as
\begin{align}
  M(x_1^n):=\min\klam{k\ge 0: -\log L_k(x_1^n)\le K(x_1^n)},
\end{align}
where $K(x_1^n)$ is the Kolmogorov complexity and $L_k(x_1^n)$ is the
maximum likelihood of order $k$,
\begin{align}
  L_k(x_1^n)
  &:=
    \max_{Q}\prod_{i=k+1}^n
    Q(x_i|x_{i-k}^{i-1})
    ,
  &
    Q(x_i|x_{i-k}^{i-1})
  &\ge 0
    ,
  &
    \sum_{x_i} Q(x_i|x_{i-k}^{i-1})
  &=1
    .
\end{align}
Function $M(x_1^n)$ is a strongly consistent and asymptotically
unbiased estimator of the Markov order. Namely, for any stationary
ergodic process $(X_i)_{i\in\mathbb{N}}$ over a finite alphabet we
have
\begin{align}
  \lim_{n\to\infty} M(X_1^n) &= M \text{ almost surely},
  &
  \lim_{n\to\infty} \mean M(X_1^n) &= M.
\end{align}
where we denote the Markov order of the process as
\begin{align}
  M:=\inf\klam{k\ge 0: P(X_{k+1}^n|X_1^k)
  =\prod_{i=k+1}^n P(X_i|X_{i-k}^{i-1})
  \text{ for all $n>k$}},
  \quad
  \inf\emptyset:=\infty.
\end{align}

Let us write succinctly the number of substrings of the supposedly
optimal length as
\begin{align}
  V(x_1^n):=V(M(x_1^n)|x_1^n).
\end{align}
Then, using the universal code by \citet{Ryabko88en2}, we can prove
inequality
\begin{align}
  \label{PreMIWords}
  J(x_1^n;x_{n+1}^{2n})
  \le
  2\kwad{D V(x_1^{2n})+
  \frac{4n\log D}{K(x_1^{2n})}+c_1}(\log n + c_2)
  ,
\end{align}
where $D$ is the cardinality of the alphabet and $c_i<\infty$ are
certain small constants \citep[Theorems 11--12]{Debowski20c}.
Applying Hilberg exponents and expectations, we obtain that the
power-law rate of mutual information is dominated by the power-law
rate of the number of words,
\begin{align}
  \label{MIWords}
  \hilberg_{n\to\infty} \mean J(X_1^n;X_{n+1}^n)
  \le
  \hilberg_{n\to\infty} \mean V(X_1^n)
  ,
\end{align}
if conditions (C) and (D) are satisfied.  The above inequality
can be chained with inequalities (\ref{CompMI}) and
(\ref{FactsComp}). The asymptotic power law
$\displaystyle\hilberg_{n\to\infty} \mean V(X_1^n)>0$ resembles
Herdan-Heaps' law for words in the linguistic sense \citep{Guiraud54,
  Herdan64, Heaps78}.

\section{Conclusion}

Chaining inequalities (\ref{CompMI}), (\ref{FactsComp}), and
(\ref{MIWords}) under conditions (A)--(D), we obtain the sandwich
bound
\begin{align}
  \label{FactsWords}
  \hilberg_{n\to\infty} \mean U_g(X_1^n;z_1^\infty)
  \le
  \hilberg_{n\to\infty} \kwad{\mean K(X_1^n)-hn}
  \le
  \hilberg_{n\to\infty} \mean J(X_1^n;X_{n+1}^n)
  \le
  \hilberg_{n\to\infty} \mean V(X_1^n),
\end{align}
which yields a formal statement of a certain theorem about facts and
words. In particular, we can infer from inequalities
(\ref{FactsWords}) that no perigraphic process can be a Markov
process, i.e., a process with a finite Markov order, $M<\infty$. These
two classes of processes are disjoint.

Of course, the above inequalities rest on a repeated use of Kolmogorov
complexity.  Therefore they are ineffective in some sense. We note
that there are other statements of theorems about facts and words that
apply effective notions. However, they are more complicated to
formulate. See also Appendix \ref{secDiscussion} in this paper for
some informal discussion of our formal assumptions.

The theorems about facts and words, as an impossibility result, raise
questions about their applicability to empirical data as well as
questions about further examples of perigraphic processes that are
more complex or more realistic than Santa Fe processes.  Some of these
questions were addressed or stated as open problems in book
\citep{Debowski21} and article \citep{Debowski21b}.


\bibliographystyle{abbrvnat}
\bibliography{0-journals-full,0-publishers-full,ai,mine,tcs,ql,books,nlp}

\appendix
\section{An informal discussion of the adopted mathematical model}
\label{secDiscussion}

A certain problem about the presented theorems about facts and words
is that they appear quite abstract. This abstraction is an advantage
from a mathematical point of view since it allows to apply the same
result to very different systems---for example to discuss the internal
complexity of communication in the field of mathematics. However, any
abstraction also begs for some concrete examples of applications that
would explain the intuitions standing behind the adopted formal
models. As a mathematician, we prefered to present the hard results in
the main matter of this paper, whereas we relegate the intuitions to
the present appendix---written upon the request of the reviewers.

First of all, one should be aware that theorems about facts and words
are an asymptotic result---for the aesthetic virtue of applying the
power law rate exponents, called succinctly Hilberg exponents. The
discussed inequalities for Hilberg exponents stem, however, from two
non-asymptotic inequalities (\ref{PreFactsComp}) and
(\ref{PreMIWords}). Thus, if we could somehow make an informed (and
necessarily fallible) guess about the Kolmogorov complexity of a
particular text in natural language or another communication system
then we might expect a sort of a theorem about facts and words also
for finite amounts of empirical data. Assumptions such as perfect
stationarity or operations like the expectation were applied to get
rid of some relatively small deviations from aesthetically appealing
general trends.

Another important remark, theorems about facts and words are just an
impossibility result: In a certain precise sense, there are always
roughly fewer facts described than words used. For Santa Fe processes,
these two quantities are of a similar order. For other processes or
communication systems, the number of facts can be significantly
smaller than the number of words. If it were so for natural language,
it would be extremely interesting. In such a case, language
communication would be much less complex than suggested by the mere
power-law growth of the lexicon. This hypothesis is so
counterintuitive that it asks for a further consideration.

From this point of view, it is advisable to bridge the adopted formal
notions of a fact and a word with a more usual understanding of these
concepts. Let us begin with the concept of a fact. As noted by one
reviewer, the representational content of a fact is irrelevant for our
theorem. This makes the theorem both powerful from a mathematical
point of view and quite difficult to digest by empirical
researchers. However, asking what the facts are in their essence is
not the most fortunate question to answer. It is more appropriate to
ask what the distinctive features of facts are in our model, namely,
how they behave.

Taking Santa Fe processes as a working mathematical model of facts,
there are three basic properties of our facts. First, they are assumed
to be binary variables, like bits, coin flips, or spins. Second, they
are independent, either probabilistically or algorithmically, namely,
their Kolmogorov complexity is the highest possible. Third, they are
persistent, recurrent, or eternal, in the sense that the same fact is
repeatedly described infinitely many times by the information source
that generates the text.

The question whether such eternal and algorithmically random facts can
be actually described at a power law rate by the totality of human
culture is intellectually challenging but important. It concerns not
only the approximate or real randomness of cultural conventions but
also their origins in the physical sources of noise. The question
cannot be honestly answered unless we know exactly how much randomness
there is in the physical world. Fortunately, in the spirit of
algorithmic information theory, we may ultimately falsify some class
of wrong answers---given enough computation time.

Moreover, it may be helpful to imagine our sequence of facts as a sort
of an unknown real parameter of an information source. When we observe
or learn this source, we learn particular binary digits or facts of
this parameter at a rate that is specific for the information
source. For Santa Fe processes, this rate is given by the power law
(\ref{Facts}). For Bernoulli processes or finite-state sources, the
same number of described facts grows only logarithmically, so the
respective Hilberg exponent is zero. By contrast, for natural
language, we cannot be sure how large it is. However---by the theorems
about facts and words---we already know that the number of independent
facts is upper bounded by the estimates of mutual information or by
the number of distinct words, which grow roughly as $n^{0.8}$, where
$n$ is the length of the text \citep{TakahiraOthers16, TanakaIshii21}.

In this way, we proceed smoothly to discuss the concept of a word.
From the point of view of formal linguistics, the only rigorous
definition of a word that makes sense is by enumeration of the
lexicon. This can be successful only if the lexicon is finite or given
by a finite formal grammar. Nevertheless, this definition cannot fully
accommodate creativity and openness of the lexicon and of other
language or culture conventions. Thus in order to count words, we need
a certain operational definition of what we want to count in general.

One such way is to count strings of letters that appear empirically
and are delimited by separators such as spaces or pauses. However,
when we count such strings in a stream of probabilistically
independent letters then we obtain a spurious power law
\citep{Mandelbrot54,Miller57}. We have a general intuition that a
memoryless mechanism cannot be responsible for the power-law
distribution of words in natural language. Thus we had better made
words operational in a different way.

For example, we may use phrases defined by universal coding. Since
there are many different universal codes of varying encoding rates for
finite data, it is not clear which one we should use. Quite an
interesting idea seems to apply grammar-based codes. These codes
represent a text as a specific context-free grammar that produces the
text as the only production. If we heuristically minimize the length
of such a grammar then we obtain non-terminals that usually span
across whole words \citep{DeMarcken96}. That is, such minimal
non-terminals can be considered a proxy for orthographic words. There
is also a theorem about facts and words that applies to the number of
minimal non-terminals \citep{Debowski11b}.  The problem is that this
theorem applies the global minimization of the grammar length, which
is likely intractable \citep{CharikarOthers05}.

For this reason, we searched for yet another approach. In the main
matter of this paper, we presented an operational proxy for words as
overlapping strings of the constant length equal to the Markov order
estimate computed for the considered text. This approach is also quite
intuitive.  Its advantage is that it does not lead to a spurious
discovery of an unbounded vocabulary in Markovian sources.  However,
the Markov order estimate is usually smaller than the average length
of a word for moderately sized texts. Thus the number of Markovian
substrings is only an imperfect proxy for the number of words.

Thus there remains an open problem of finding an operational
definition of a word that could be applied to any sort of symbolic
data and would possess the following properties:
\begin{enumerate}
\item The number of distinct operational words should be theoretically
  lower bounded by the block mutual information---and hence by the
  number of facts. (Thus, a rich semantics implies a rich vocabulary.)
\item The number of distinct operational words should be theoretically
  upper bounded for a finite Markov order of the process.  (Thus, a
  meager semantics implies a meager vocabulary.)
\item Parsing of the input text into operational words should be
  efficiently computable in a time close to linear in the length of
  the text.
\item The operational words should be similar in shape and in number
  to the orthographic words for natural language data.
\end{enumerate}
We hope that such a satisfactory operational definition of a word
exists.

Concluding this appendix, there seem to be a few somewhat different
formal statements that fall under the umbrella of theorems about facts
and words. In our work, we tried to identify a few of them but the
topic has not been exhausted, in our opinion.

\end{document}